\documentstyle[aps,epsf]{revtex}

\begin{document}

\preprint{CUMQ/HEP 103}

\title{The Supersymmetric Spectrum in $SO(10)$ GUTs with Gauge-Mediated
Supersymmetry Breaking}

\author{M.~Frank$^{\text{a}}$, H.~Hamidian$^{\text{b}}$,
K.~Puolam\"{a}ki$^{\text{c}}$}

\address{$^{\text{a}}$Department of Physics, Concordia University, Montreal,
Quebec, Canada H3G 1M8 }
\address{$^{\text{b}}$Department of Physics, University of Illinois at Chicago,
Chicago, Illinois, 60607-7059 }
\address{$^{\text{c}}$Helsinki Institute of Physics, P.O.Box 9,
FIN-00014 University of Helsinki, Finland }

\date{June 16, 1999}

\maketitle
\begin{abstract}
 We investigate a large class of supersymmetric $SO(10)$
grand unified theories within the framework of gauge-mediated supersymmetry
breaking.
 We start with the
most general messenger sector and imbedd the standard model gauge group
into either
$SU(2)_L \times SU(2)_R
\times U(1)_{B-L}$ or $SU(2)_L \times U(1)_{I_{3R}}\times U(1)_{B-L}$. We
find that
the conditions
of the perturbativity of the gauge couplings and unification at the GUT
scale severely restrict the messenger sector and lead to testable
phenomenological predictions for the sparticle masses. One of the most notable
features of the class of supersymmetric $SO(10)$ grand unified theories
introduced here is that among the many possible cases there are only a few
consistent models, all of which share an essentially unique mass spectrum.
\end{abstract}

\section{Introduction}

Supersymmetry (SUSY) is currently believed to lead to the
most attractive scenario of physics beyond the standard model (SM).
However, the low-energy spectrum of fermions and bosons
does not exhibit this symmetry, so SUSY, if it exists, must somehow
be broken to be of any relevance to Nature.
SUSY breaking is different from the more familiar symmetry breaking
in the SM since
the supertrace theorem prevents SUSY breaking by
tree-level renormalizable couplings. According to this theorem one must assume
that the sector
responsible for SUSY breaking is hidden, i.e. it has no
renormalizable tree-level couplings with the visible sector. A popular
realization of this breaking can be accommodated through gravity,
a theory which is altogether non-renormalizable, and this mechanism  has
been explored
for a long time \cite{acn}.
Another possibility, in which a different mechanism is used to
communicate SUSY breaking,
can be provided by keeping the original theory
renormalizable, but with a low-energy description in terms of an effective
Lagrangian
with non-renormalizable terms. Within the first scenario (i.e. theories with
gravity-mediated SUSY breaking) soft mass
terms are generated at the Planck scale and one cannot produce
flavor-invariant SUSY-breaking terms for the sfermion masses.
On the other hand, in the second scenario it is possible to generate soft
terms
at some ``messenger''
scale $\Lambda_M$ below
the Planck scale and to break flavor symmetry only through Yukawa
couplings. This can be
achieved within the framework of theories which take advantage of the
general mechanism
of gauge-mediated supersymmetry breaking (GMSB) \cite{gmsb} in which,
just as in the SM, the Yukawa couplings are the only sources of flavor
violation.
In GMSB theories one introduces
vector-like quarks and leptons whose role is to break supersymmetry.
These theories have received considerable attention
in recent years
\cite{gmsbnew}, largely
due to their high predictability and their emphasis on a dynamical origin for
SUSY breaking.

To carry out the GMSB program, one starts with an
observable sector which contains the usual matter and gauge fields and
their supersymmetric partners, while leaving the hidden sector
unspecified.
One then introduces a {\it messenger} sector, formed by the new
superfield $X$, whose
coupling with the goldstino superfield generates a supersymmetric mass of
order
$S$ for the messenger fields, and leads to mass splittings of order $F$
(where we denote by $S$, $F$ the vacuum expectation values of the scalar and
the auxiliary component of the superfield, respectively), and thus
$\sqrt{F}$ is
identified with the scale of SUSY breaking in the messenger sector
\cite{dineetal}.
Apart from the requirement that the messenger fields transform under the SM
gauge group,
the messenger sector is unknown and is the main
source of model-dependence in GMSB theories.

Explicit model building in theories which take advantage of the GMSB mechanism
has been explored in a number
of papers by taking $SU(5)$ to be the unifying gauge group in most cases
\cite{gmsbph,borzu,skiba,martin,chanowitz}.
Despite serious
shortcomings related to problems such as the nucleon decay rates and
neutrino masses, as
well as $R$-parity breaking, the predictability and simplicity of these
models
has served as a useful first illustration of the applicability and power of
the idea of GMSB.
An attempt to go
beyond these models was made by the authors of
\cite{mohapatra}. They chose to study the electroweak gauge group
$SU(2)_L\times
U(1)_{I_{3 R}}
\times U(1)_{B-L}$ since it automatically guarantees $R$-parity
conservation. They also explored messenger fields which can play the role
of the Higgs fields and break the chosen gauge group down to the SM.
In this work we propose to study the GMSB mechanism and its
phenomenological consequences
in a large class of supersymmetric
$SO(10)$ grand unified theories (GUTs) by taking a different approach.  We
start
with the most general messenger sector and break the $SO(10)$ symmetry to
either
$SU(2)_L\times U(1)_{I_{3 R}} \times U(1)_{B-L}$ or $SU(2)_L\times
SU(2)_R \times U(1)_{B-L}$, with a subsequent breaking to the SM gauge
group. We enforce
the boundary condition that the gauge couplings must unify at a common scale
by using the one-loop renormalization group evolution of the couplings.
We use this boundary condition to restrict the messenger field masses and
to predict
the complete sparticle mass spectrum.

Our paper is organized as follows. In section II, we
describe the $SO(10)$ model and its breaking. In section III we
introduce the
$SU(2)_L\times U(1)_{I_{3 R}} \times U(1)_{B-L}$ and $SU(2)_L\times
SU(2)_R \times U(1)_{B-L}$ models and give their particle content. In
section IV we
discuss the restrictions on SUSY breaking and present the resulting
solutions. The particle spectrum is discussed in section V.
We conclude with some remarks in section VI.

\section{$SO(10)$ models}

$SO(10)$ GUTs have received a great deal of attention over the years and
continue to do so,
thanks to the improved
measurements of the low-energy gauge couplings which confirm that
supersymmetry leads to an extremely accurate (perturbative) unification of
couplings
 \cite{bb}.
Also, from a more theoretical point of view,
the $SO(10)$ gauge group is a natural candidate for supersymmetric
unification since all the quarks and leptons of a single generation are
the components of a single spinor representation. Furthermore, the
$SO(10)$ gauge group has the particularly attractive feature
that it can break to the SM through either
$U(1)_{I_{3 R}}$ or $SU(2)_R$, thus providing non-zero neutrino masses through
the see-saw mechanism. Due to this, and in view of the recent observations
at SuperKamiokande
\cite{Kamiokande}
which indicate non-vanishing neutrino masses, $SO(10)$ GUTs
must be considered even more seriously than before.
Additionally, $SO(10)$ GUTs also provide interesting fermion mass
relations.
For example, depending on the scale of the right-handed symmetry breaking,
one can predict lower bounds on neutrino masses.
In this context, it was shown that $SO(10)$ is a realistic
and natural supersymmetric grand-unified theory \cite{bb}. Yet another
 important feature of $SO(10)$ SUSY GUTs is that they make it possible
to achieve the desired doublet-triplet splitting (i.e. keeping the pair
of Higgs doublets of the supersymmetric SM light, while giving
their colour triplet partners superheavy masses to avoid proton decay)
without fine-tuning \cite{dtsp}.

Let us now turn our attention to $SO(10)$ symmetry breaking in $SO(10)$
SUSY GUTs.
$SO(10)$ symmetry breaking can proceed in essentially two ways:
$SO(10)$ can be broken to $SU(5) \times U(1)$ at scale $\sim 10^{18}$GeV
and then further broken down to the minimal supersymmetric standard model
(MSSM) at scale $\sim
10^{16}$GeV. Or, alternatively, $SO(10)$ can be broken to some
left-right  symmetric group, e.g. $G_{224}=SU(2)_L\times SU(2)_R\times
SU(4)_C$, at some high energy scale, where the $SU(4)_C$ group contains the
subgroup $SU(3)_C\times U(1)_{B-L}$. $G_{224}$ is then broken at some
intermediate scale to the MSSM. The latter possibility of breaking
$SO(10)$ is of particular interest, since the supersymmetric left-right
model can naturally account for parity and comes with the extra bonus of
providing possible solutions to both the SUSY CP problem and the
R-parity problem \cite{mohras}. In this work we will further assume the
D-terms to vanish (they could in principle contribute to the
supersymmetry breaking masses). It can be argued that renormalizable
see-saw mechanism and spontaneously broken $B-L$ symmetry lead to
exact $R$-parity at all energies \cite{amrs}.  These are important and
attractive features of the left-right models, which is why we wish to
study this possibility of breaking through the GMSB mechanism in more
detail in this paper.

The symmetry breaking chain that we study here is as follows.
We assume that the $SO(10)$ gauge group is
broken to an intermediate left-right symmetry group $G_{LR}$ at scale
$M_G$. $M_G$ should be no less than $10^{16}$~GeV to have stable
nuclei and it should be below the Planck scale $10^{19}$~GeV. The
intermediate symmetry group $G_{LR}$ is then broken down to the MSSM at scale
$M_R$. For simplicity, and to maintain only a minimal number of scales in the
theory, we will assume that there are no further symmetry breaking
scales between $M_R$ and $M_G$.

As possible left-right symmetry groups we will consider $G_{LR}^I=SU(3)_C
\times SU(2)_L \times U(1)_{I_{3 R}} \times U(1)_{B-L}$ and
$G_{LR}^{II}=SU(3)_C \times SU(2)_L \times SU(2)_R \times U(1)_{B-L}$.
Next we must include SUSY breaking into our model.

The simplest possibility which avoids the proliferation of scales is to assume
that the SUSY breaking scale and the left-right gauge
symmetry breaking scale could be somehow related.
To realize this we will further assume that the
SUSY breaking scale $\Lambda_{\text{SUSY}}$ is
the same as the left-right symmetry breaking scale $M_R$. In terms of the
scales
introduced in the previous section, we relate
explicitly the scales as $\Lambda_{\text{SUSY}}=\frac{F}{S}$.
This is an attractive choice since it simultaneously connects the scale of
the gauge
symmetry breaking to the scale of SUSY breaking and fulfills the
requirement that the
breaking of the electroweak  symmetry remain radiative.In order for the
sparticle
masses to be around 1 TeV, the SUSY breaking scale must be
$\Lambda_{\text{SUSY}}
\sim M_R
\sim 100$~TeV.
Our specification of scales leads, among other things, to an interesting
scenario
which involves a low
right-handed scale, effects of which could be observed in precision
measurements of low energy observables at LHC or NLC \cite{dmm}.

To sum up, in the symmetry breaking chains studied here we restrict the
scales as follows:
\begin{equation}
10^5{\text{GeV}}<M_R<10^7{\text{GeV}} {\text{ and }}
10^{16}{\text{GeV}}<M_G<10^{19}{\text{GeV}} .
\label{eq:scales}
\end{equation}

\noindent
We would like to note that this particular choice of scales is
in agreement with general astrophysical/cosmological bounds
derived from nucleosynthesis \cite{ggr}.

In GMSB models it is the messenger fields that carry the information about
SUSY breaking to the visible sector.  In our case the messenger
sector is strongly limited by the constraint that the gauge couplings
must meet at scale
$M_G$ and that they remain perturbative up to the Planck
scale. We will study these constraints and show that among all the possible
choices for
messenger-field contents and multiplicities only a
few models remain which are consistent with all the constraints that we
impose.
(In view of the fact that we start out with a minimal number of scales and
assumptions,
this is a somewhat intriguing result and
we are inclined to take these phenomenological implications to be very
suggestive.)
Once the messenger sector is specified, one can calculate all the SUSY
breaking parameters of the MSSM and thus make predictions for the complete
mass spectrum in the MSSM. As described below, we have fully carried out
this program by taking the one-loop radiative effects into account.

\section{Left-right models}

As has been mentioned above, we assume that the $SO(10)$ symmetry breaking
proceeds
through one of the intermediate left-right (product) gauge groups
$SU(2)_L\times U(1)_{I_{3R}} \times U(1)_{B-L}$ or
$SU(2)_L\times SU(2)_R \times U(1)_{B-L}$, followed
by a subsequent breaking down to the SM.
Let us now briefly describe the superpotentials and the particle contents in
theories with these gauge groups.

\subsection*{Model~I: $SU(2)_L\times U(1)_{I_{3R}} \times U(1)_{B-L}$}

The model based on the gauge group $SU(2)_L\times U(1)_{I_{3R}} \times
U(1)_{B-L}$
is phenomenologically interesting because it contains all the
usual matter multiplets {\it plus} the right-handed neutrinos. Furthermore,
this model forbids lepton- and baryon-number violating terms in the
superpotential, thus guaranteeing automatic $R$-parity
conservation \cite{mohras}. The model can therefore lead to a stable lightest
supersymmetric  particle (LSP) which can be a cold dark matter (CDM)
candidate. The
superpotential for the  matter sector of the theory is:
\begin{eqnarray}
\label{superpotential1}
W & = &  h_{u} Q H_u u^{c} + h_d Q H_d d^{c} + h_e L H_d e^{c}
+ h_{\nu} L H_u \nu^{c} \nonumber \\
& & + \mu H_u H_d + f \delta \nu^c \nu^c + M_R \delta {\bar \delta}+ W_m,
\end{eqnarray}
where $W_m$ denotes the messenger sector superpotential. The lepton and the
quark sectors
in this theory consists of the doublets $Q(2,0,1/3)$,
$L(2,0,-1)$, and the singlets $u^c(1,-1/2, -1/3)$, $d^c(1, 1/2,
-1/3)$, $e^c(1, 1/2, 1)$, $\nu^c(1, -1/3, 1)$, and the
corresponding squarks and sleptons. The two Higgs doublets (and their
superpartners)
in this model are the same as in the MSSM:
$H_u(2, 1/2, 0)$ and
$H_d( 2, -1/2, 0)$. In addition to these, the model contains two $SU(2)_L$
Higgs
singlets $\delta (1, 1, -2)$ and ${\bar \delta} (1, -1, 2)$ which break the
$U(1)_{I_{3R}} \times U(1)_{B-L}$ symmetry down to the $U(1)_Y$ of the
SM, together with their superpartners. The gauge sector of the the model
contains the bosons
$W(3, 0, 0)$,~$B(1, 0, 0)$ and $V(1, 0, 0)$, and the corresponding
gauginos.

\subsection*{Model II: $SU(2)_L\times SU(2)_R \times U(1)_{B-L}$}

The left-right supersymmetric model based on the gauge group $SU(2)_L\times
SU(2)_R \times
U(1)_{B-L}$ has been studied intensively over the years \cite{fh}.
In addition to the attractive features of Model~I, this model offers
solutions to both the strong and weak CP
problems, while at the same time preserving $R$-parity.
As opposed to the $SU(2)_L \times U(1)_{I_{3R}}$ model,
in the $SU(2)_L \times SU(2)_R$ model
$R$-parity conservation is not automatic. However, there exist ways to avoid
it being broken spontaneously \cite{mohras,amrs}. The superpotential for
the matter
sector of this theory is:

\begin{eqnarray}
\label{superpotential2}
W & = & {\bf h}_{q}^{(i)} Q_L^T\tau_{2}\Phi_{i} \tau_{2}Q_R +
{\bf h}_{l}^{(i)}
L_L^T\tau_{2}\Phi_{i} \tau_{2}L_R + i({\bf h}_{LR}L_L^T\tau_{2} \delta_L
L_L + {\bf h}_{LR}L_R^{T}\tau_{2}
\Delta_R L_R) \nonumber \\
& & + M_{LR}\left[ Tr(\Delta_L {\bar \delta}_L + Tr(\Delta_R {\bar
\delta}_R)\right] + \mu_{ij}Tr(\tau_{2}\Phi^{T}_{i} \tau_{2} \Phi_{j})+W_m,
\end{eqnarray}
where, as before, $W_m$ denotes the messenger sector superpotential.
The particle content in this theory is as follows. The lepton and the quark
sectors consist of the doublets $Q_L(2,1,1/3)$,
~$L_L(2,1,-1)$,~$ Q_R(1, 2, -1/3)$,~$L_R( 1, 2 ,1)$, and the corresponding
superpartners. This model contains bi-doublet Higgs fields $\Phi_u (2, 2,
0)$ and
$\Phi_d( 2, 2, 0)$, the triplet Higgs fields
$\Delta_L (3,1,-2)$ and $\Delta_R (1,3,-2)$
which break the
left-right model to the SM, as well as $\delta_L (3,1,2)$,
$\delta_R (0,3,2)$ (which are required for the cancellation of anomalies in the
fermionic sector), and their superpartners. The gauge sector consists of
the bosons
$W_L(3, 1, 0), W_R(1, 3, 0), B(1, 0, 0)$, and $V(1, 1, 0)$ and their
associated superpartners.

\section{The messenger sector}

The simplest messenger sector consists of $N_f$ flavors of chiral
superfields $\Phi_i$ and ${\bar \Phi}_i$, $(i=1,\cdots, N_f)$ which
transform in the ${\bf r} +{\bf {\bar r}}$ representation of the gauge
group. In
order to preserve gauge coupling constant unification, one usually
requires that the messengers form complete GUT multiplets. In this case
the presence of the messenger fields at an intermediate scale does not
modify the value of $M_G$. The gauge coupling strength receives an
extra contribution from the messengers, at the unification scale
$M_G$ which is given by \cite{dineetal}:
\begin{eqnarray}
\delta \alpha^{-1}_{GUT}=-\frac{N}{2 \pi}\ln\frac{M_G}{S},
\end{eqnarray}
with
\begin{eqnarray}
N=\sum^{N_f}_{i=1} n_i.
\end{eqnarray}
Here $n_i$ is twice the Dynkin index of the ${\bf r}$ representation of the
gauge group
and $i$ is the flavor index. By requiring that the gauge interactions remain
perturbative all the way up to the GUT scale $M_G$ one obtains:
\begin{eqnarray}
N \alt 150/\ln \frac{M_G}{S}.
\end{eqnarray}

On the one hand, the minimal set of messenger fields can be relaxed and
augmented further if one requires a theory sufficiently rich to be
viable---for instance by requiring a reasonable set of masses for the
scalars and gauginos. On the other hand, the number of messenger
fields is restricted by the requirement that the MSSM couplings stay
perturbative up to the GUT scale and that gauge couplings unify to a
single coupling at $M_G$.  The requirement of gauge unification
does not provide much information on the messenger sector: indeed it
is not necessary that all messenger-scale vector-like superfields
obtain their masses by coupling primarily to the chiral superfield
$X$. The messengers that do not satisfy this condition have little
effect on the MSSM masses, but can still participate in gauge
coupling unification. The possible messenger fields in Model~I are:

\begin{eqnarray}
Q_8 = (8, 1, 0, 0) , \nonumber \\
L_3 = (1, 3, 0, 0) , \nonumber \\
\Delta+\overline \Delta = (1, 3 , 0, -2) +conj. , \nonumber \\
\Delta^c +\overline {\Delta^c} = (1, 1 , -1, 2) +conj. , \nonumber \\
H + \overline H =  (1, 2, \frac{1}{2}, 0) + conj.  , \nonumber \\
Q + \overline Q = (3, 2, 0, \frac{1}{3}) + conj.  ,\nonumber \\
U ^c+\overline{U^c} =({\overline 3}, 1, - \frac{1}{2}, -\frac{1}{3})
+conj. , \nonumber\\
D ^c+\overline{D^c}=({\overline 3}, 1 ,\frac{1}{2},- \frac{1}{3})
+conj. , \nonumber\\
L +\overline L = (1, 2, 0,  -1) +conj. , \nonumber\\
e^c +\overline{e^c} = (1, 1, \frac {1}{2} , 1) +conj. , \nonumber  \\
\nu^c +\overline{\nu^c}= ( 1, 1, -\frac {1}{2}, 1) +conj. ,
\end{eqnarray}
and they transform under $SU(3)_C \times SU(2)_L \times U(1)_{I_{3 R}}
\times U(1)_{B-L}$ as specified by the quantum numbers in brackets.

Similarly, the possible messenger fields in Model~II consist of
\begin{eqnarray}
Q_8=(8,1,1,0) , \nonumber \\
Q_3=(1,3,1,0) , \nonumber \\
Q_3^c=(1,1,3,0) , \nonumber \\
\phi=(1,2,2,0) , \nonumber \\
Q +\overline Q=(3,2,1,\frac {1}{3})+conj. , \nonumber \\
Q^c +\overline {Q^c}=(3,1,2,-\frac {1}{3})+conj. , \nonumber \\
L+\overline L=(1,2,1,-1)+conj. , \nonumber \\
L^c+\overline{L^c}=(1,1,2,1)+conj. , \nonumber \\
\Delta +\overline \Delta =(1,3,1,-2) + conj. , \nonumber \\
\Delta^c+\overline {\Delta^c} =(1,1,3,2) + conj.,
\end{eqnarray}
which transform under $SU(3)_C\times
SU(2)_L \times SU(2)_R \times U(1)_{B-L}$ with the quantum
numbers specified in brackets.

In the following section we shall restrict the messenger sectors by
requiring (i) the perturbativity of the gauge couplings ($\alpha_k<1$)
up to the Planck scale and (ii) unification  at the GUT scale $M_G$.

\section{The SO(10) solution}

Unification is the requirement that the values of the four gauge couplings
be equal
to a single value $\alpha_G$ at scale $M_G$. We use subindex
``$V$'', as opposed to ``$B-L$'', to denote the GUT-normalized $B-L$
gauge coupling. The relation between the $U(1)_{B-L}$ gauge
coupling $\alpha_{B-L}$ (analogous to the hypercharge $\alpha_Y$ of
the SM) and the GUT-normalized gauge coupling $\alpha_V$
(analogous to the $\alpha_1$ of the standard model) is $\alpha_V=
\frac{2}{3}~\alpha_{B-L}$. At the left-right breaking scale
$M_R$ we match couplings to those of the the MSSM. Denoting by
$\beta_L$, $\beta_R$, $\beta_{V}$ and $\beta_C$ the
beta-functions corresponding to the $SU(2)_L$, $SU(2)_R$ (or
$U(1)_{I_{3R}}$), $U(1)_{V}$,
and $SU(3)_C$ gauge groups respectively,
  the one-loop
renormalization group equations at the $M_R$ scale are as follows:
\begin{eqnarray}
\alpha_L^{-1}(M_R) = \alpha_G^{-1} +\beta_L (t_G-t_R) , \nonumber \\
\alpha_R^{-1}(M_R) = \alpha_G^{-1} +\beta_R (t_G-t_R) , \nonumber \\
\alpha_{V}^{-1}(M_R) = \alpha_G^{-1} +\beta_{V} (t_G-t_R) , \nonumber \\
\alpha_C^{-1}(M_R) = \alpha_G^{-1} +\beta_C (t_G-t_R) ,
\label{eq:rng1}
\end{eqnarray}
where we have defined:
\begin{equation}
t_R = \frac 1{2 \pi} \ln \frac{M_R}{M_Z} ~~{\text{ and }}~~ t_G =
\frac 1{2 \pi} \ln \frac{M_G}{M_Z}.
\end{equation}
The one-loop matching conditions to the MSSM at the scale $M_R$ are:
\begin{eqnarray}
\frac 53 \alpha_1^{-1}(M_R)&=&\alpha_R^{-1}(M_R)+\frac 23
\alpha_{V}^{-1}(M_R) , \nonumber \\
\alpha_2^{-1}(M_R)&=&\alpha_L^{-1}(M_R) , \nonumber \\
\alpha_3^{-1}(M_R)&=&\alpha_C^{-1}(M_R),
\label{eq:rng2}
\end{eqnarray}
where $\alpha_1$, $\alpha_2$ and $\alpha_3$ are the gauge couplings of
$U(1)_Y$, $SU(2)_L$ and $SU(3)_C$ respectively.
Combining equations (\ref{eq:rng1}) and (\ref{eq:rng2}) yields
\begin{equation}
\alpha_k^{-1}(M_Z)=\alpha_G^{-1}+\beta_k^{MSSM} t_R+\beta_k^{LR}
(t_G-t
_R) , k=1,2,3 ,
\label{eq:rng3}
\end{equation}
where we have
defined the left-right $\beta$-functions by
\begin{equation}
\beta^{LR}=\left( \begin{array}{c} \beta_1^{LR} \\
\beta_2^{LR} \\ \beta_3^{LR}
\end{array} \right) =\left(  \begin{array}{c}
\frac 35 \beta_R+\frac
25 \beta_{V} \\ \beta_L \\ \beta_C \end{array}
\right) .
\label{eq:definebeta}
\end{equation}
and where the
MSSM $\beta$-functions
$\beta_k^{MSSM}=(33/5,1,-3)^T$ have been used.  We can eliminate the
unification
coupling $\alpha_G$ from equation (\ref{eq:rng3})
by using
constraints from equation (\ref{eq:scales}). We thus obtain the
following
limits on the differences of LR $\beta$-functions:
\begin{equation}
3.1 < \beta_2^{LR}-\beta_3^{LR} < 4.1 {\text{
and }} 7.4 <
\beta_1^{LR}-\beta_3^{LR} < 9.9 .
\label{eq:constrain1}
\end{equation}

The choice of messenger fields is determined by the requirement that the
gauge couplings remain perturbative ($\alpha_k<1$)
up to the Planck scale, as well as by the requirement of obtaining
plausible masses for all MSSM particles.
The requirement of the perturbativity of the gauge couplings at the scale
$M_G$ and between $M_G$ and $M_{P}$,
together with (\ref{eq:rng3}), leads to the following constraints on the
left-right $\beta$-functions:
\begin{equation}
\beta_1^{LR}<10.4 {\text{ , }} \beta_2^{LR}<6.1
{\text{ and }}
\beta_3^{LR}<3.0 .
\label{eq:constrain2}
\end{equation}
These equations constrain the number of messenger fields,
since each new chiral superfield contributes to the $\beta$-functions.

The requirement that the gauge couplings unify at a specific energy
scale may not always impose severe restrictions on the messenger
sector, but it certainly provides for reasonable MSSM masses and agrees
well with the
apparent experimental confirmation for gauge coupling unification at LEP.
Using the $\beta$-functions for
Models I and II (see Appendix C), together with the constraints
(\ref{eq:constrain1})
and (\ref{eq:constrain2}), we find that there can be
no consistent solutions in
Model II. In Model~I, on the other hand, there exist consistent
solutions with the messenger
multiplicities:
\begin{equation}
n_8=n_3=n_H+n_L=1 {\text{ and }}
n_{e^c}+n_{\nu^c}=0,1 .
\label{eq:messconstrain}
\end{equation}
We will
study these possibilities in more detail and show that despite apparently
having several choices, their consequences are remarkably
alike for all the models specified by
(\ref{eq:messconstrain}).

According to (\ref{eq:messconstrain}) the messenger sector consists of
one color octet ($n_8=1$) field, one $SU(2)_L$ triplet ($n_3=1$)
field, and a pair of $H$- or $L$-type (Higgs- or Lepton-like, respectively)
messenger fields
($n_H+n_L=1$). There could also be a pair of $e^c$ or $\nu^c$ type fields
($n_{e^c}+n_{\nu^c}=0,1$). There are thus a total of six choices for
the messenger multiplicities. However, each of these messenger sectors
produces a very similar mass spectrum for the MSSM,
making our scheme extremely predictive. We have listed the spectrum
obtained from these models in Table~I (for
$\Lambda_{SUSY}=100~TeV, \Lambda_M=100~\Lambda_{SUSY}$) and Table~II
(for $\Lambda_{SUSY}=50~TeV,~\Lambda_M=10~\Lambda_{SUSY}$). In both
Tables the value of the bilinear scalar coupling ${\tilde B}_{\mu}$ is kept
fixed and
$\tan \beta$ is allowed to vary. Note that in the second case only one
model survives and
the stau mass is barely positive. In the first Table the solutions
for 5 out of 6 scenarios are obtained. As can be seen from these Tables, both
of the representative cases require a large $\tan \beta \approx 35-40$.

In every case studied the $SU(2)_L$
and the $SU(3)_C$ gauge couplings meet at $M_G=2.0 \times 10^{16}$GeV
(which we take to be our GUT scale) for
$10^5{\text{GeV}}<M_R<10^7{\text{GeV}}$.
For the choice $n_{e^c}+n_{\nu^c}=0$ the coupling $\alpha_1$
corresponding
to the $U(1)_{I_{3R}} \times U(1)_V$ gauge groups and
$\alpha_3$
corresponding to the color gauge group meet at $M_G'=1.9
\times
10^{16}$GeV, or within 6\% of the GUT scale.
For the choice $n_{e^c}+n_{\nu^c}=1$
the mismatch is much worse: $\alpha_1$ meets the
colour gauge coupling
at $M_G' \simeq 4-5 \times 10^{15}$GeV. However, this
mismatch can be accounted for and explained through the threshold effects
and so we will consider
this model as well.
Finally, in left-right models the
ratio:
\begin{equation}
\tan^2
\theta_R=\frac{\alpha_{B-L}(M_R)}{\alpha_R(M_R)},
\end{equation}
is a very important phenomenological parameter which is completely
determined by
the unification condition. Here the value of $\tan^2 \theta_R$ lies
in the interval $1.3 \le \tan^2 \theta_R \le 1.6$.

The solutions presented above do not include the fields
$\Delta+\overline{\Delta}
=(1,3,0,-2) +conj.$ and
$\Delta^c+\overline {\Delta^c} =(1,1,-1, 2)+conj.$ at the MSSM scale. These
fields are
essential for  the see-saw mechanism and conservation of R-parity. The
contribution
of these fields is included in the renormalization-group equations between
$M_G$ and
$M_R$. Including them in the  interval between $M_{MSSM}=10^3$ GeV and
$M_R=10^5$ GeV would affect only slightly the $\beta_1$ function,
translating in a change of the messenger scale
by a factor of 1.5, from $10^5-10^7$ GeV to $1.5 \times 10^5-10^7$ GeV,
which is insignificant compared to the
accuracy we are working with \cite{chacko}.

\section{The sparticle spectrum}

Given a messenger sector, and once the messenger scale and $\tan^2 \theta_R$
are fixed, one
can calculate the sparticle masses in the corresponding left-right
model (see appendix A). The left-right model is then matched to the
MSSM (appendix B). Knowing the MSSM parameters at the messenger scale, we
can then
calculate the full MSSM particle spectrum as a function of $\tan
\beta$~\cite{hamidian}. One could reduce the parameters
further and solve the supersymmetric $CP$-problem, as was done
in~\cite{kaiandotherfriends}, by requiring that the bilinear scalar
coupling ${\tilde B}_\mu$ vanish at the messenger scale. This would fix $\tan
\beta$. However, in this work we wish to take a broader course
by treating $\tan \beta$ (or equivalently the bilinear scalar coupling) as
a free parameter. We list representative results for the sparticle
spectrum obtained for the six models by choosing $\tan \beta = 15$ in
Table~III and $\tan \beta = 2$ in Table~IV.

>From Table~\ref{tab:beta1}, where we list the spectrum in terms of
$\Lambda_{\text SUSY}$, $\Lambda_M=\lambda S$ (the messenger scale:
see appendix A), and ${\tilde B}_\mu$, one can see that the squarks
 are quite heavy in this model. The heavy top squark drives the
radiative symmetry breaking and we find that for all values of the
parameters the mass-squared term of the up-type Higgs boson indeed
acquires a negative value, thus always leading to the desired
(radiatively-induced) symmetry breaking.

\begin{figure}
\caption{Masses of the light spartners in six allowed models.}
\label{fig:mass6}
\mbox{\epsfxsize=16cm \epsfysize=16cm \epsffile{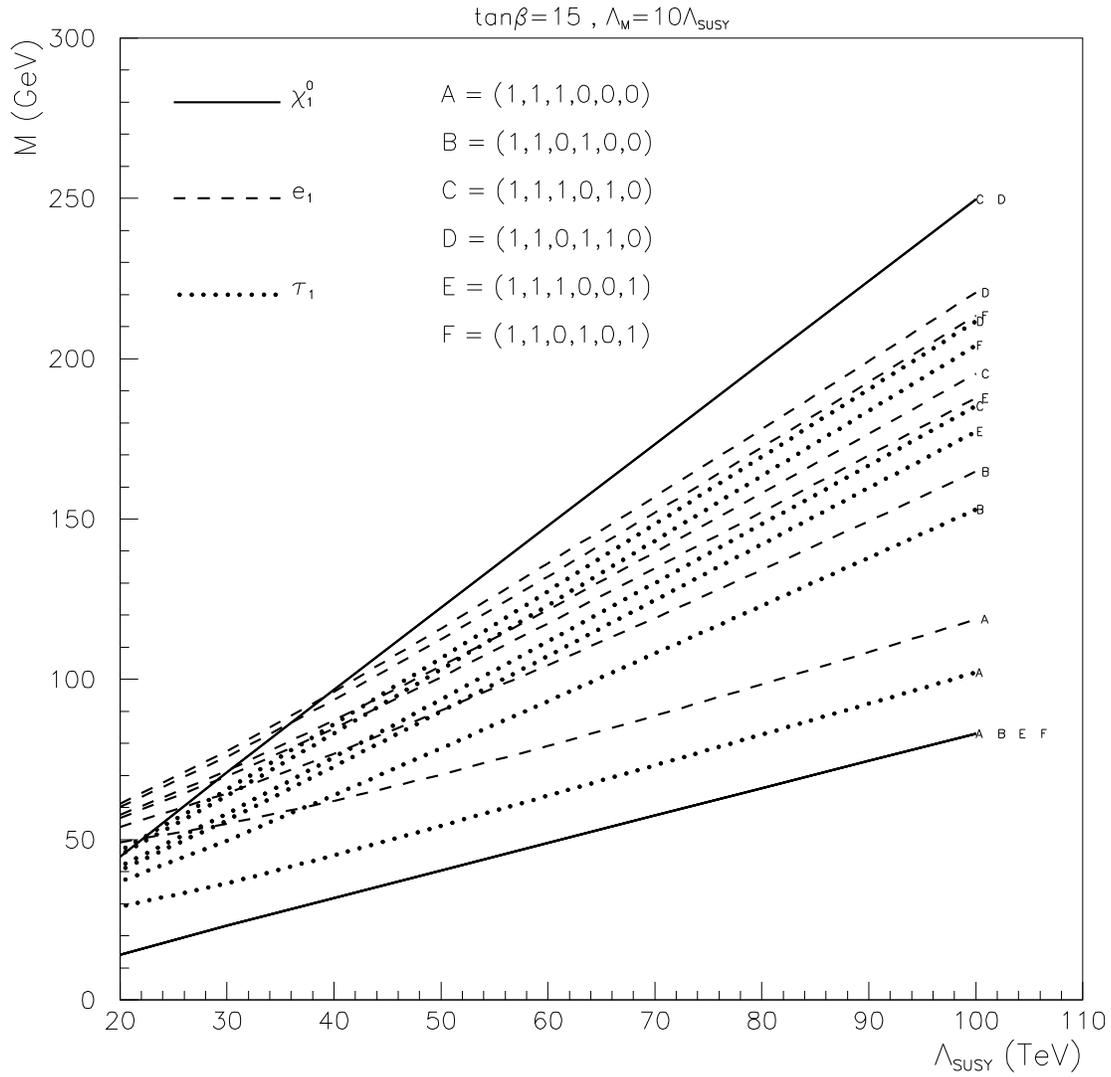}}
\end{figure}

As the gauge symmetry is always radiatively broken, one must check
that the resulting vacuum is a physical one, i.e. that all
mass-squared eigenvalues of the charged scalars remain positive and
above the current experimental limits. Here, as in other GMSB models, the
lightest scalar fermion turns out to be the lighter stau mass
eigenstate. For low values of $\Lambda_{\text{SUSY}}$ the lighter stau
has a mass below the experimental limit of about $72$~GeV
\cite{LEP2}. In Figure~\ref{fig:mass6} we have plotted the masses of the
lighter
sparticles as a function of scale
$\Lambda_{\text{SUSY}}$.
One can see that the stau is always light and the lower bound on the
stau mass sets a lower limit on the scale of supersymmetry breaking and
the sparticle masses. We have plotted this limit in
Figure~\ref{fig:limit3} as a function of $\tan
\beta$. It is seen that the squark mass scale must always be heavy, of
the order of $\sim$~1~TeV. At high values
of $\tan \beta$ ($\tan \beta
\agt 30$) one would require multi-TeV squarks in order to satisfy the LEP2
exclusion limit on stau mass.

Despite the fact that our restrictions allow six different solutions,
we find that all of them are somewhat remarkably similar in predictions for
the supersymmetric spectrum.
The characteristic features of the obtained mass spectrum for the
supersymmetric
partners (and
$H^{\pm}$) are:
\begin{itemize}

\item Depending on the exact messenger content, the next-to-lightest
supersymmetric particle (NLSP) can be
either the lighter stau or the lightest neutralino, which is
essentially a gaugino. The choice of $n_{e^c}=1$ favours the
neutralino as the NLSP, while solutions with $n_{e^c}=0$ favour the
stau as the NLSP. One can notice, from eq. (C4) and (C5), that in the
latter case, the Bino is the NLSP. If $n_{e^c}=1$, the Bino mass is a
triplet. One has approximately the folowing ratios for the gaugino masses
($M_k\propto \alpha_kN_k$ and the 1,2,3 stand for $U(1)_Y,~SU(2)_L$ and
$SU(3)_C$
 respectively):
$$n_{e^c}=0~~~~M_1:M_2:M_3=0.029:0.29:1$$
$$n_{e^c}=1~~~~M_1:M_2:M_3=0.088:0.29:1$$

\item As expected, the lighter selectron is always heavier than the lighter
stau.
Also, the scalar leptons are always lighter than the
squarks, which turn out to be very heavy in this model. From
Figure~\ref{fig:limit3}, one can see that the theory always predicts a
light stau and, equivalently, heavy squarks ($\agt 1$~TeV for small $\tan
\beta$ and
$\agt 2$~TeV for large $\tan \beta$).  Note that the usual hierarchy
$m_{{\tilde e}_{1,2}}
\leq m_{{\tilde d}_{1,2}}
\approx m_{{\tilde u}_{1,2}} \leq m_{{\tilde t}_{1,2}}$ holds. In this
model, $m_{{\tilde l}_{1,2}}$ and $m_{{\tilde q}_{1,2}}$ are
mixed states of left and right sleptons or squarks.

\item The sneutrinos are always heavier than the lighter of the
charged sleptons, unlike in supersymmetric models without GMSB. In
fact $m_{{\tilde \nu}_{e,\tau}} \approx m_{{\tilde \tau}_2}$. The charged
sleptons are lighter than the Higgs/Higgsinos. (The masses
of the right-handed sleptons are around 100~GeV and the masses of the
left-handed
sleptons around 300 GeV for minimal squark masses.)

\item The bilinear Higgs coupling, the so-called $\mu$ parameter in the
superpotential, lies in the $400-500$~GeV
region and can be {\em either} positive {\em or} negative, as shown in
Table~\ref{tab:modellist}, as opposed to the values obtained in
\cite{nandi}. As expected in GMSB theories,
$\mu^2 > m_{{\tilde l}_{1,2}}^2$. Here $\mu
\cong M_{\tilde q}/100$, meaning that
$|{\mu}|$ is always at least 50~GeV. This makes the Higgsino at least
twice as heavy as the wino, so that the light charginos are mostly gauginos.
If we tune the value of the bilinear scalar coupling (the
${\tilde B}_{\mu}$ parameter) to zero at the messenger scale, we obtain a
positive value for $\mu$. In general, the sign of $\mu$ does not seem
to make much difference to the mass spectrum. However, we obtain sizable
effects
in, for example, the $b \rightarrow s\gamma$ decay width, since for
positive $\mu$ the
interference between the SM and chargino contributions is destructive,
whereas for negative $\mu$ it is constructive.

\item The heavy spartner masses are quite accurately directly
proportional to the scale $\Lambda_{\text{SUSY}}= F/S$. We have listed the
particle content for all our six models in
Table~\ref{tab:modellist}. For the choice of $\tan \beta=15$ and
$\Lambda_{\text{SUSY}}=50$ TeV the squark masses are around $1$~TeV,
while the charged Higgs boson mass is about $0.5$~TeV in all cases. The
heavy Higgs mass is of the order of the $\mu$-term. The heavy sleptons,
neutralinos and charginos all have masses between
$310$~GeV and $470$~GeV.

\item Finally, if one includes a supergravity sector, one would
expect a light (mass in the eV range) gravitino. As in other GMSB theories,
we
assume the gravitino to be the LSP.

\end{itemize}

\section{Summary and Conclusion}
If the low-energy world of the standard model indeed descends from a
supersymmetric
theory, there could be a plethora of experimental signals at future
facilities and
the results of theoretical analyses will be needed
to distinguish among a large variety of supersymmetric
scenarios. Most of the parameters in SUSY are associated with
the SUSY breaking sector. Within the framework of perturbative unification,
one starts with a SUSY GUT scenario, specifies the scale
and mechanism for SUSY breaking,
and imposes phenomenological constraints to reduce the number of free
parameters (as many as 124 in the MSSM only! \cite{h}). Here we have chosen
to study supersymmetric $SO(10)$ GUTs with GMSB as
viable SUSY GUT candidates, since these are free from the problems that
plague $SU(5)$-based theories, with or without GMSB. By breaking $SO(10)$
to left-right symmetric gauge groups, and by taking advantage of the
extreme predictive power of the GMSB mechanism, we calculate and discuss a
number
of important phenomenological results. As in any other SUSY GUT, with GMSB or
otherwise, a number of conditions must be imposed before a detailed study
is made.
We have kept the number of such (boundary) conditions to a minimum to
examine {\em all}
such SUSY GUTs with GMSB in a large class of $SO(10)$ theories.
The conditions imposed were that the gauge couplings remain perturbative
($\alpha_k<1$) up to the Planck scale, and that they unify at the
$GUT$ scale. Except for the left-right symmetry breaking and the messenger
scales,
no other intermediate scales are assumed, and we take the SUSY and the
left-right
breaking scales, $\Lambda_{\text{SUSY}}$ and $M_R$ respectively, to be the
same.
 From the condition that sparticle masses be of the order of $\sim 1$~TeV,
the SUSY breaking scale must be $\Lambda_{\text{SUSY}} \sim 100$~TeV.
Remarkably, these conditions rule out a large number of scenarios, and
the ones that survive exhibit surprisingly similar features. The NLSP is
either the neutralino (with a large gaugino component) or the lighter
stau, with masses of about 50~GeV.
This would be testable either at LEP~II through direct production
of neutralinos
\cite{lnz}, or at the Tevatron/LHC through the production of two lepton jets
plus missing energy \cite{nandi}. However, one could see that the chargino
mass is above 300 GeV, which is just below the experimental limit
in $b \rightarrow s \gamma$ decay. This would make it more likely to
observe a signal at
the LHC, whereas we expect LEP to be able to observe at most the stau
or the lightest Higgs boson. Future collider experiments can put all
these predictions fully to test.

Finally, let us note that although here we have studied
$SO(10)$ GUTs equipped with the GMSB mechanism as the
next logical choice beyond similar theories based on the $SU(5)$
gauge group, the supersymmetric spectrum obtained is not unlike the one
obtained in gauge-mediated models with a simpler breaking chain. This
raises the attractive possibility that
gauge-mediated breaking imposes similar features on a variety of GUT
scenarios. It would certainly be
interesting to study other GUT scenarios with GMSB to test what appear to
be general features of this
mechanism.

\section*{Aknowledgements}

We thank Z. Chacko, M. Raidal, A. Riotto and T. Kobayashi for comments and
for raising interesting questions.
The work of M.F. was partially funded by
NSERC of Canada (SAP0105354).  The work of H.H. is supported by the
U.S. Department of Energy under Grant Number DE-FG02-91ER40676 and
that of K.P. partially by the Academy of Finland (No.~37599).

\appendix

\section{Supersymmetry breaking mass terms in
theories containing
single adjoint messenger fields}
\label{sec:adjoint}

The superpotential can be written
as:
\begin{equation}
W_m=\sum_i \lambda_i X \Phi_i {\overline{\Phi}}_i +\sum_j
\frac 12
\lambda_j X {\rm Tr}(Q_j^2-Q_j'^2),
\end{equation}
where $\Phi_i+{\overline{\Phi}}_i$ are the
usual messenger fields and $Q$ and $Q'$ are additional
messenger fields transforming in the
adjoint representation of the gauge
group. $X$ is the chiral singlet superfield which
parametrizes supersymmetry breaking, $<X>=S+\theta^2 F$, and is related to the
messenger scale through the explicit relation $\Lambda_M=\lambda S$, and
$\lambda_{i,j}$ are Yukawa couplings. The gauge interactions of $Q$ and
$Q'$ written in terms of superfields
in the interaction Lagrangian have the form:
\begin{equation}
Q^\dagger_j T^a Q_j+Q'^\dagger_j T^a Q'_j {\text{ and }}
Q^\dagger_j T^a T^b
Q_j+Q'^\dagger_j T^a T^b Q'_j .
\end{equation}
The one-loop
diagrams contributing to the gaugino mass terms and the
two-loop diagrams
contributing to the scalar mass-squared terms
contain either field $Q_j$ or
field $Q'_j$, but never both of them at the
same time.  The contributions
from fields $Q_j$ and $Q'_j$ are thus
separate and equal.

Next define new
superfields $V_j$ and ${\overline{V}}_j$ by:
\begin{equation}
\left(
\begin{array}{c} V_j \\ {\overline{V}}_j \end{array} \right) =
\frac
1{\sqrt{2}} \left( \begin{array}{cc} 1 & 1 \\ -1 & 1 \end{array}
\right)
\left( \begin{array}{c} Q_j \\ Q'_j \end{array}
\right) .
\end{equation}
The fields
$V$ and ${\overline{V}}$ transform under
gauge transformations defined by $V  \rightarrow (1-i\epsilon_a T^a)V$.

The
contribution of the $\lambda X \Phi_i{\overline \Phi}_i$ part of the
superpotential
to the sparticle mass terms is known
~\cite{dineetal}. Since the contributions of the adjoint fields
$Q_j$ and $Q'_j$
to the gaugino and sparticle mass terms is, to leading
order, of equal
magnitude, we can isolate the contribution of a single
adjoint
field.

The  gaugino
masses are generated at one loop, where the particles in the loop are the
messenger fields \cite{martin}:
\begin{equation}
M_a=\frac{\alpha_a}{4\pi} \frac FS \sum_i
n_a(i) g(x_i), a=1,2,3,
\end{equation}
and the mass-squared terms for the
scalars are generated from two-loop diagrams, with messenger fields, gauge
bosons, and gauginos as the internal lines \cite{martin}:
\begin{equation}
{\tilde{m}}^2_k=2 \left| \frac FS \right| ^2
\sum_a \left(
\frac{\alpha_a}{4\pi} \right)^2 C_a(k) \sum_i n_a(i) f(x_i),
\label{eq:masssquared}
\end{equation}
where $n_a(i)$ denotes the sum of the Dynkin indices for the messenger pair
$\Phi_i$ and
${\overline{\Phi}}_i$. For example, for $N+{\overline{N}}$ of $SU(N)$
$n_{SU(N)}(N+{\overline{N}})=1$. In the case
of a single adjoint messenger field $Q$, $n_a(Q)$ is the Dynkin index of
that
particular field. For example, for a single $SU(N)$ adjoint
messenger field $Q$ one has
$n_{SU(N)}(Q)=N$. In the GUT
normalized
$Q_V+{\overline Q}_V = (B-L)+{\overline{(B-L)}}$ of $U(1)_{B-L}$,
$n_{V}((B-L)+{\overline{(B-L)}})=3(B-L)^2$, and for $Q_R+{\overline Q}_R
=R+{\overline{R}}$ of
$U(1)_{I_{3 R}}$ one has $n_R(R+{\overline{R}})=2R^2$.

In the formula~(\ref{eq:masssquared}) $C_a(k)$ is the quadratic
Casimir invariant of the scalar field $k$ in
question, normalized such that $C_a=4/3$ for $SU(3)$ triplets,
$C_2=3/4$ for $SU(2)$ doublets,
$C_V=\frac 32 (B-L)^2$ for
$U(1)_{B-L}$ fields, and $C_R=R^2$ for $U(1)_R$
fields.

Finally, as in \cite{martin}, $x$, $f(x)$, and $g(x)$ in the
above formulas are defined as:
\begin{eqnarray}
x_i&&=\left| \frac F{\lambda_i S^2} \right| ,
\nonumber \\
g(x)&&=\frac 1{x^2} \left[ (1+x) \ln (1+x)+(1-x) \ln(1-x)
\right] ,
\nonumber \\
f(x)&&=\frac{1+x}{x^2} \left[ \ln(1+x)-2
{\text{Li}}_2\left( \frac
x{1+x} \right) +\frac 12 {\text{Li}}_2 \left(
\frac {2x}{1+x} \right)
\right] + (x \rightarrow -x ).
\end{eqnarray}
Functions $g(x)$ and $f(x)$ can be taken to equal unity to a good
approximation.

\section{Scalar and gaugino masses
at the messenger scale for specific fields}

The gaugino mass terms can
be written as
\begin{equation}
M_a= \frac{\alpha_a}{4 \pi}  \frac FS N_a ,
\end{equation}
where $N_a$ denote the sum of the Dynkin indices for each messenger field::
\begin{equation}
\left( \begin{array}{c} N_L \\ N_R \\ N_V \\ N_C \end{array} \right) =
\left( \begin{array}{c} 2  \\ 0 \\ 0 \\ 0 \end{array} \right) n_3
+\left( \begin{array}{c} 0 \\ 0 \\ 0 \\ 3 \end{array} \right) n_8
+\left( \begin{array}{c} 1 \\ 1 \\ 0 \\ 0 \end{array} \right) n_H
+\left( \begin{array}{c} 1 \\ 0 \\ \frac 32 \\ 0 \end{array} \right) n_L
+\left( \begin{array}{c} 0 \\ \frac 12 \\ \frac 34 \\ 0 \end{array} \right)
(n_{e^c}+n_{\nu^c}) .
\end{equation}

When the LR model is matched to the MSSM the MSSM gaugino masses are
obtained most easily from the formula:
\begin{equation}
M_a=\frac{\alpha_a}{4
\pi} \frac FS  N_a', a=1,2,3
,
\end{equation}
where:
\begin{equation}
N'= \left( \begin{array}{c} 0 \\ 2
\\ 0 \end{array} \right) n_3
   +\left( \begin{array}{c} 0 \\ 0 \\ 3
\end{array} \right) n_8
   +\left( \begin{array}{c} \frac 35 \\ 1 \\ 0
\end{array} \right) (n_H+ n_L)
   +\left( \begin{array}{c} \frac 65 \\ 0 \\ 0
\end{array} \right) n_{e^c}
   +\left( \begin{array}{c} 0 \\ 0 \\ 0
\end{array} \right) n_{\nu^c}
.
\end{equation}

The scalar mass-squared terms can be
expressed as the sum:
\begin{equation}
{\tilde{m}}^2_k \simeq 2 \sum_{a=1}^4
C_a(k) \left| \frac FS \right| ^2
\left(
\frac {g_a}{4 \pi} \right)^4 N_a,
\end{equation}
and by matching the LR model to the MSSM the
MSSM scalar mass-squared terms are easily calculated. The
$SU(2)_L$ and $SU(3)_C$ matchings also proceed in a straightforward manner.

\section{The $\beta$-functions}

Defining the $\beta$-functions according to:
\begin{equation}
\frac d{dt} \alpha_k^{-1}=\frac 1{2 \pi} \beta_k,
\end{equation}
we now give the explicit expressions for the $\beta$-functions in models I
and II.

The $\beta$-functions for Model I are:
\begin{eqnarray}
\left(
\begin{array}{c} \beta_L \\ \beta_R \\ \beta_V \\ \beta_C
\end{array}
\right)& = &
\left( \begin{array}{c} -6 \\ 0 \\ 0 \\ -9 \end{array} \right)
+
\left( \begin{array}{c} 2 \\ 2 \\ 2 \\ 2 \end{array} \right) n_g +
\left(
\begin{array}{c} 1 \\ 1 \\ 0 \\ 0 \end{array} \right) n_H +
\left(
\begin{array}{c} 0 \\ 2 \\ 3 \\ 0 \end{array} \right) n_{\Delta^c}+
\left(
\begin{array}{c} 4 \\ 0 \\ 9 \\ 0 \end{array} \right) n_\Delta
 +
\left( \begin{array}{c} 0 \\ 0 \\ 0 \\ 3 \end{array} \right) n_8
\nonumber \\
&+&
\left( \begin{array}{c} 2 \\ 0 \\ 0 \\ 0 \end{array} \right) n_3 +
\left(
\begin{array}{c} 3 \\ 0 \\ \frac 12 \\ 2 \end{array} \right)
n_Q +
\left(
\begin{array}{c} 0 \\ \frac 32 \\ \frac 16 \\ 1 \end{array}
\right)
(n_{u^c}+n_{d^c}) +
\left( \begin{array}{c} 1 \\ 0 \\ \frac 32 \\ 0
\end{array} \right)
n_L +
\left( \begin{array}{c} 0 \\ \frac 12 \\ \frac 34
\\ 0 \end{array}
\right) (n_{e^c}+n_{\nu^c}) .
\end{eqnarray}
The minimal
model without any messengers has three generations
($n_g=3$) and one pair of
Higgs doublets and Higgs triplets
($n_H=n_{\Delta^c}=1$).

The
$\beta$-functions for Model II are:
\begin{eqnarray}
\left(
\begin{array}{c} \beta_L \\ \beta_R \\ \beta_V \\ \beta_C
\end{array}
\right) & = &
\left( \begin{array}{c} -6 \\ -6 \\ 0 \\ -9 \end{array} \right)
+
\left( \begin{array}{c} 2 \\ 2 \\ 2 \\ 2 \end{array} \right) n_g +
\left(
\begin{array}{c} 1 \\ 1 \\ 0 \\ 0 \end{array} \right) n_\phi +
\left(
\begin{array}{c} 0 \\ 4 \\ 9 \\ 0 \end{array} \right) n_{\Delta^c}+
\left(
\begin{array}{c} 4 \\ 0 \\ 9 \\ 0 \end{array} \right) n_\Delta
 +
\left( \begin{array}{c} 0 \\ 0 \\ 0 \\ 3 \end{array} \right) n_8
\nonumber \\
& +&
\left( \begin{array}{c} 2 \\ 0 \\ 0 \\ 0 \end{array} \right) n_3 +
\left(
\begin{array}{c} 0 \\ 2 \\ 0 \\ 0 \end{array} \right) n_{3^c} +
\left(
\begin{array}{c} 3 \\ 0 \\ \frac 12 \\ 2 \end{array} \right)
n_Q +
\left(
\begin{array}{c} 0 \\ 3 \\ \frac 12 \\ 2 \end{array} \right)
n_{Q^c}
+
\left( \begin{array}{c} 1 \\ 0 \\ \frac 32 \\ 0 \end{array} \right)
n_L
+
\left( \begin{array}{c} 0 \\ 1 \\ \frac 32 \\ 0 \end{array}
\right)
n_{L^c} .
\end{eqnarray}
The minimal model without any messengers
consist of three generations
($n_g=3$), two Higgs bidoublets ($n_\phi=2$), and
one pair of
right-handed Higgs triplets ($n_{\Delta^c}=1$).

The values of
$\beta$-functions $\beta^{LR}_k$, $k=1,2,3$, are listed
in
Tables~\ref{tab:beta1} and~\ref{tab:beta2}.


%

\begin{table}
\caption{$\Lambda_{\text{SUSY}}=100$TeV,
$\Lambda_M=100\Lambda_{\text{SUSY}}$, $B_
M=0$.}
\begin{tabular}{lrrrrrrr}
$(n_3,n_8,n_
H,n_L,n_{e^c},n_{\nu^c})$ & $\tan \beta$ & $\mu$ & $m_{H^\pm}$
&
$m_{{\tilde{\chi}}^\pm_{1,2}}$ & $m_{{\tilde{\chi}}^0_{1,2,3,4}}$
&
$m_{{\tilde{e}}_{1,2}}$ & $m_{{\tilde{\tau}}_{1,2}}$ \\ & $\frac{\Gamma
(b
\rightarrow s \gamma)}{\Gamma_{SM}}$ & $M_3$
&
$m_{{\tilde{\nu}}_e}/m_{{\tilde{\nu}}_{\tau}}$ & $m_{{\tilde{u}}_{1,2}}$
&
$m_{{\tilde{t}}_{1,2}}$ & $m_{{\tilde{d}}_{1,2}}$
&
$m_{{\tilde{b}}_{1,2}}$ \\
\tableline $(1,1,0,1,0,0)$ & $37$ &
$913$ &
 $912$ &
 $746/909$ &
 $83/746/887/908$ & $172/661$ &
 $34/659$ \\

& $1.1$ & $2001$ & $656/648$ & $1887/1992$ &
 $1663/1871$ & $1887/1993$ &
$1815/1869$ \\
\tableline $(1,1,1,0,1,0)$ & $38$ & $905$ &
 $905$
&
 $745/902$ &
 $250/745/879/902$ & $207/653$ &
 $109/652$ \\
 & $1.1$ &
$2001$ & $648/640$ & $1892/1991$ &
 $1668/1868$ & $1892/1993$ & $1816/1867$
\\
\tableline $(1,1,0,1,1,0)$ & $38$ & $910$ &
 $904$ &

$746/906$ &
 $250/746/884/906$ & $234/671$ &
 $152/670$ \\
 & $1.1$ &
$2001$ & $667/658$ & $1890/1992$ &
 $1665/1870$ & $1890/1993$ & $1814/1868$
\\
\tableline $(1,1,1,0,0,1)$ & $37$ & $906$ &
 $913$ &

$745/902$ &
 $83/745/879/902$ & $197/652$ &
 $95/651$ \\
 & $1.1$ & $2001$
& $648/640$ & $1892/1991$ &
 $1667/1870$ & $1892/1993$ & $1819/1869$ \\
\tableline $(1,1,0,1,0,1)$ & $37$ & $910$ &
 $912$ &
 $746/907$ &

$83/746/884/906$ & $224/670$ &
 $141/668$ \\
 & $1.1$ & $2001$ & $666/657$
& $1889/1992$ &
 $1665/1871$ & $1890/1993$ & $1817/1869$
\\
\end{tabular}
\end{table}

\begin{table}
\caption{$\Lambda_{\text{SUSY}}=50$T
eV,
$\Lambda_M=10\Lambda_{\text{SUSY}}$, ${\tilde B}_\mu = 0$.}
\begin{tabular}{lrrrrrrr}
$(n_3,n_8,n_
H,n_L,n_{e^c},n_{\nu^c})$ & $\tan \beta$ & $\mu$ & $m_{H^\pm}$
&
$m_{{\tilde{\chi}}^\pm_{1,2}}$ & $m_{{\tilde{\chi}}^0_{1,2,3,4}}$
&
$m_{{\tilde{e}}_{1,2}}$ & $m_{{\tilde{\tau}}_{1,2}}$ \\ & $\frac{\Gamma
(b
\rightarrow s \gamma)}{\Gamma_{SM}}$ & $M_3$
&
$m_{{\tilde{\nu}}_e}/m_{{\tilde{\nu}}_{\tau}}$ & $m_{{\tilde{u}}_{1,2}}$
&
$m_{{\tilde{t}}_{1,2}}$ & $m_{{\tilde{d}}_{1,2}}$
&
$m_{{\tilde{b}}_{1,2}}$ \\
\tableline $(1,1,0,1,1,0)$ & $41$ & $438$ &

$422$ &
 $348/466$ &
 $122/348/425/466$ & $115/327$ &
 $3/337$ \\
 & $1.2$
& $1060$ & $317/314$ & $1000/1045$ &
 $905/1007$ & $1001/1048$ & $956/1000$
\\
\end{tabular}
\end{table}

\begin{table}
\caption{$\Lambda_{\text{SUSY}}=50$T
eV,
$\Lambda_M=10\Lambda_{\text{SUSY}}$, fixed $\tan
\beta=15$.}
\label{tab:modellist}
\begin{tabular}{lrrrrrrr}
$(n_3,n_8,n_H,n_L,n_
{e^c},n_{\nu^c})$ & $\tan \beta$ & $\mu$ & $m_{H^\pm}$
&
$m_{{\tilde{\chi}}^\pm_{1,2}}$ & $m_{{\tilde{\chi}}^0_{1,2,3,4}}$
&
$m_{{\tilde{e}}_{1,2}}$ & $m_{{\tilde{\tau}}_{1,2}}$ \\ & $\frac{\Gamma
(b
\rightarrow s \gamma)}{\Gamma_{SM}}$ & $M_3$
&
$m_{{\tilde{\nu}}_e}/m_{{\tilde{\nu}}_{\tau}}$ & $m_{{\tilde{u}}_{1,2}}$
&
$m_{{\tilde{t}}_{1,2}}$ & $m_{{\tilde{d}}_{1,2}}$
&
$m_{{\tilde{b}}_{1,2}}$ \\
\tableline $(1,1,1,0,0,0)$ & $15$ & $439$ &
 $514$ &
 $346/469$ &
 $40/346/426/469$ & $70/312$ &
 $54/314$ \\
 & $1.2$ & $1060$ & $302/302$ & $1000/1045$ &
 $905/1019$ & $1001/1048$ & $990/1006$ \\
\tableline $(1,1,0,1,0,0)$ & $15$ & $441$ &
 $513$ &
 $347/471$ &
 $40/347/429/470$ & $90/322$ &
 $78/323$ \\
 & $1.2$ & $1060$ & $312/312$ & $999/1045$ &
 $904/1019$ & $1000/1048$ & $989/1006$ \\
\tableline $(1,1,1,0,1,0)$ & $15$ & $438$ &
 $514$ &
 $345/468$ &
 $122/346/425/468$ & $104/318$ &
 $93/320$ \\
 & $1.2$ & $1060$ & $309/308$ & $1001/1045$ &
 $906/1019$ & $1002/1048$ & $990/1007$ \\
\tableline $(1,1,0,1,1,0)$ & $15$ & $440$ &
 $514$ &
 $346/470$ &
 $122/347/427/470$ & $115/327$ &
 $106/328$ \\
 & $1.2$ & $1060$ & $317/317$ & $1000/1045$ &
 $905/1019$ & $1001/1048$ & $990/1007$ \\
\tableline $(1,1,1,0,0,1)$ & $15$ & $438$ &
 $514$ &
 $345/468$ &
 $40/346/425/468$ & $100/318$ &
 $89/320$ \\
 & $1.2$ & $1060$ & $308/308$ & $1001/1045$ &
 $906/1019$ & $1002/1048$ & $990/1007$ \\
\tableline $(1,1,0,1,0,1)$ & $15$ & $440$ &
 $513$ &
 $346/470$ &
 $40/346/428/470$ & $112/326$ &
 $103/328$ \\
 & $1.2$ & $1060$ & $317/316$ & $1000/1045$ &
 $905/1019$ & $1001/1048$ & $990/1007$ \\
\tableline $(1,1,1,0,0,0)$ & $15$ & $-439$ &
 $510$ &
 $354/463$ &
 $41/353/428/462$ & $70/312$ &
 $54/314$ \\
 & $1.3$ & $1060$ & $302/302$ & $1000/1045$ &
 $908/1016$ & $1001/1048$ & $989/1005$ \\
\tableline $(1,1,0,1,0,0)$ & $15$ & $-441$ &
 $509$ &
 $355/465$ &
 $41/354/431/463$ & $90/322$ &
 $78/323$ \\
 & $1.3$ & $1060$ & $312/312$ & $999/1045$ &
 $907/1016$ & $1000/1048$ & $988/1005$ \\
\tableline $(1,1,1,0,1,0)$ & $15$ & $-438$ &
 $511$ &
 $354/462$ &
 $123/353/426/461$ & $104/318$ &
 $93/320$ \\
 & $1.3$ & $1060$ & $309/308$ & $1001/1045$ &
 $909/1016$ & $1002/1048$ & $989/1006$ \\
\tableline $(1,1,0,1,1,0)$ & $15$ & $-440$ &
 $510$ &
 $354/464$ &
 $123/354/429/463$ & $115/327$ &
 $106/328$ \\
 & $1.3$ & $1060$ & $317/317$ & $1000/1045$ &
 $908/1016$ & $1001/1048$ & $989/1006$ \\
\tableline $(1,1,1,0,0,1)$ & $15$ & $-438$ &
 $511$ &
 $354/462$ &
 $41/353/427/461$ & $100/318$ &
 $90/320$ \\
 & $1.3$ & $1060$ & $308/308$ & $1001/1045$ &
 $909/1016$ & $1002/1048$ & $989/1006$ \\
\tableline $(1,1,0,1,0,1)$ & $15$ & $-440$ &
 $510$ &
 $354/464$ &
 $41/354/429/462$ & $112/326$ &
 $103/328$ \\
 & $1.3$ & $1060$ & $317/316$ & $1000/1045$ &
 $908/1016$ & $1001/1048$ & $989/1006$ \\
\end{tabular}
\end{table}

\begin{table}
\caption{$\Lambda_{\text{SUSY}}=50$T eV,
$\Lambda_M=10\Lambda_{\text{SUSY}}$, fixed $\tan \beta=2$.}
\label{tab:models2}
\begin{tabular}{l|l|rrrrrrrrrrr}
$(n_3,n_8,n_H,n_L,n_
{e^c},n_{\nu^c})$ & $\tan \beta$ & $\mu$ & $m_{H^\pm}$
&
$m_{{\tilde{\chi}}^\pm_{1,2}}$ & $m_{{\tilde{\chi}}^0_{1,2,3,4}}$
&
$m_{{\tilde{e}}_{1,2}}$ & $m_{{\tilde{\tau}}_{1,2}}$ \\ & $\frac{\Gamma
(b
\rightarrow s \gamma)}{\Gamma_{SM}}$ & $M_3$
&
$m_{{\tilde{\nu}}_e}/m_{{\tilde{\nu}}_{\tau}}$ & $m_{{\tilde{u}}_{1,2}}$
&
$m_{{\tilde{t}}_{1,2}}$ & $m_{{\tilde{d}}_{1,2}}$
&
$m_{{\tilde{b}}_{1,2}}$ \\
\tableline $(1,1,1,0,0,0)$ & $2$ & $622$ &
 $758$ &
 $364/624$ &
 $38/365/600/626$ & $64/311$ &
 $64/311$ \\
 & $0[[1.]]$ & $1060$ & $305/305$ & $1001/1045$ &
 $867/1022$ & $1001/1047$ & $990/1001$ \\
\tableline $(1,1,0,1,0,0)$ & $2$ & $624$ &
 $757$ &
 $364/625$ &
 $38/365/602/627$ & $85/321$ &
 $85/321$ \\
 & $0[[1.]]$ & $1060$ & $315/315$ & $999/1046$ &
 $867/1022$ & $1000/1047$ & $990/1000$ \\
\tableline $(1,1,1,0,1,0)$ & $2$ & $622$ &
 $759$ &
 $364/623$ &
 $120/365/600/626$ & $100/317$ &
 $99/317$ \\
 & $0[[1.]]$ & $1060$ & $311/311$ & $1002/1046$ &
 $868/1022$ & $1002/1047$ & $990/1002$ \\
\tableline $(1,1,0,1,1,0)$ & $2$ & $623$ &
 $758$ &
 $364/624$ &
 $120/365/601/627$ & $112/325$ &
 $112/325$ \\
 & $0[[1.]]$ & $1060$ & $320/320$ & $1001/1046$ &
 $867/1022$ & $1001/1048$ & $991/1001$ \\
\tableline $(1,1,1,0,0,1)$ & $2$ & $622$ &
 $759$ &
 $364/623$ &
 $38/364/600/626$ & $96/317$ &
 $96/317$ \\
 & $0[[1.]]$ & $1060$ & $311/311$ & $1001/1046$ &
 $868/1022$ & $1002/1047$ & $990/1002$ \\
\tableline $(1,1,0,1,0,1)$ & $2$ & $623$ &
 $758$ &
 $364/625$ &
 $38/365/601/627$ & $108/325$ &
 $108/325$ \\
 & $0[[1.]]$ & $1060$ & $319/319$ & $1000/1046$ &
 $867/1022$ & $1001/1048$ & $991/1001$ \\
\end{tabular}
\end{table}

\begin{table}
\caption{Contributions of
different messenger multiplicities to the
beta functions $\beta_k^{LR}$,
$k=1,2,3$, (see
eq. (8)) in Model I. $\beta_0^{LR}$ is the
contribution of
the minimal model.}
\label{tab:beta1}
\begin{tabular}{l|l|rrrrrrrrr}

&
Limit & $\beta_0^{LR}$ & $n_H+n_L$ & $n_{\Delta^c}$ & $n_\Delta$ & $n_8$
&
$n_3$ & $n_Q$ & $n_{u^c}+n_{d^c}$ & $n_{e^c}+n_{\nu^c}$ \\
\tableline

$\beta_1^{LR}$ & $<10.4$ & $9$ & $\frac 35$ & $\frac {12}5$ & $\frac {18}5$
& $0$ & $0$ & $\frac 15$ & $\frac{29}{30}$ & $\frac 35$ \\ \tableline

$\beta_2^{LR}$ & $<6.1$ & $1$ & $1$ & $0$ & $4$ & $0$ & $2$ & $3$ & $0$ &
$0$ \\ \tableline

$\beta_3^{LR}$ & $<3.0$ & $-3$ & $0$ & $0$ & $0$ & $3$ & $0$ & $2$ & $1$ &
$0$ \\ \tableline

$\beta_2^{LR}-\beta_3^{LR}$ & $3.1...4.1$ & $4$ & $1$ & $0$ & $4$ & $-3$ &
$2$ &
$1$ & $-1$ & $0$ \\ \tableline

$\beta_1^{LR}-\beta_3^{LR}$ & $7.4...9.4$ & $12$ & $\frac 35$ & $\frac{12}5$ &
$\frac{18}5$ & $-3$ & $0$ & $-\frac 95$ & $-\frac 1{30}$ & $\frac 35$

\end{tabular}
\end{table}

\begin{table}
\caption{Contributions of different messenger multiplicities to the
beta functions $\beta_k^{LR}$, $k=1,2,3$, (see
eq. (8)) in Model II. $\beta_0^{LR}$ is the
contribution of the minimal model.}
\label{tab:beta2}
\begin{tabular}{l|l|rrrrrrrrrrr}

& Limit & $\beta_0^{LR}$ & $n_\phi$ & $n_{\Delta^c}$ & $n_\Delta$ &
$n_8$ & $n_3$ & $n_{3^c}$ & $n_Q$ & $n_{Q^c}$ & $n_L$ & $n_{L^c}$ \\
\tableline

$\beta_1^{LR}$ & $<10.4$ & $9.6$ & $\frac 35$ & $6$ & $\frac {18}5$ &
$0$ & $0$ & $\frac 65$ & $\frac 15$ & $2$ & $\frac 35$ & $\frac 65$ \\
\tableline

$\beta_2^{LR}$ & $<6.1$ & $2$ & $1$ & $0$ & $4$ & $0$ & $2$ & $0$ & $3$ &
$0$ & $1$ & $0$ \\ \tableline

$\beta_3^{LR}$ & $<3.0$ & $-3$ & $0$ & $0$ & $0$ & $3$ & $0$ & $0$ & $2$ &
$2$ & $0$ & $0$ \\ \tableline

$\beta_2^{LR}-\beta_3^{LR}$ & $3.1...4.1$ & $-5$ & $1$ & $0$ & $4$ &
$-3$ & $2$ & $0$ & $1$ & $-2$ & $1$ & $0$ \\ \tableline

$\beta_1^{LR}-\beta_3^{LR}$ & $7.4...9.4$ & $12.6$ & $\frac 35$ & $6$ &
$\frac{18}5$ & $-3$ & $0$ & $\frac 65$ & $-\frac 85$ & $0$ & $\frac
35$ & $\frac 65$

\end{tabular}
\end{table}

\begin{figure}
\caption{Lower limit on squark masses as a function of $\tan \beta$ in
six allowed model. The limit has been obtained by requiring the
lighter stau to have a mass of at least 72GeV.}
\label{fig:limit3}
\mbox{\epsfxsize=16cm \epsfysize=16cm \epsffile{kuva32.epsi}}
\end{figure}

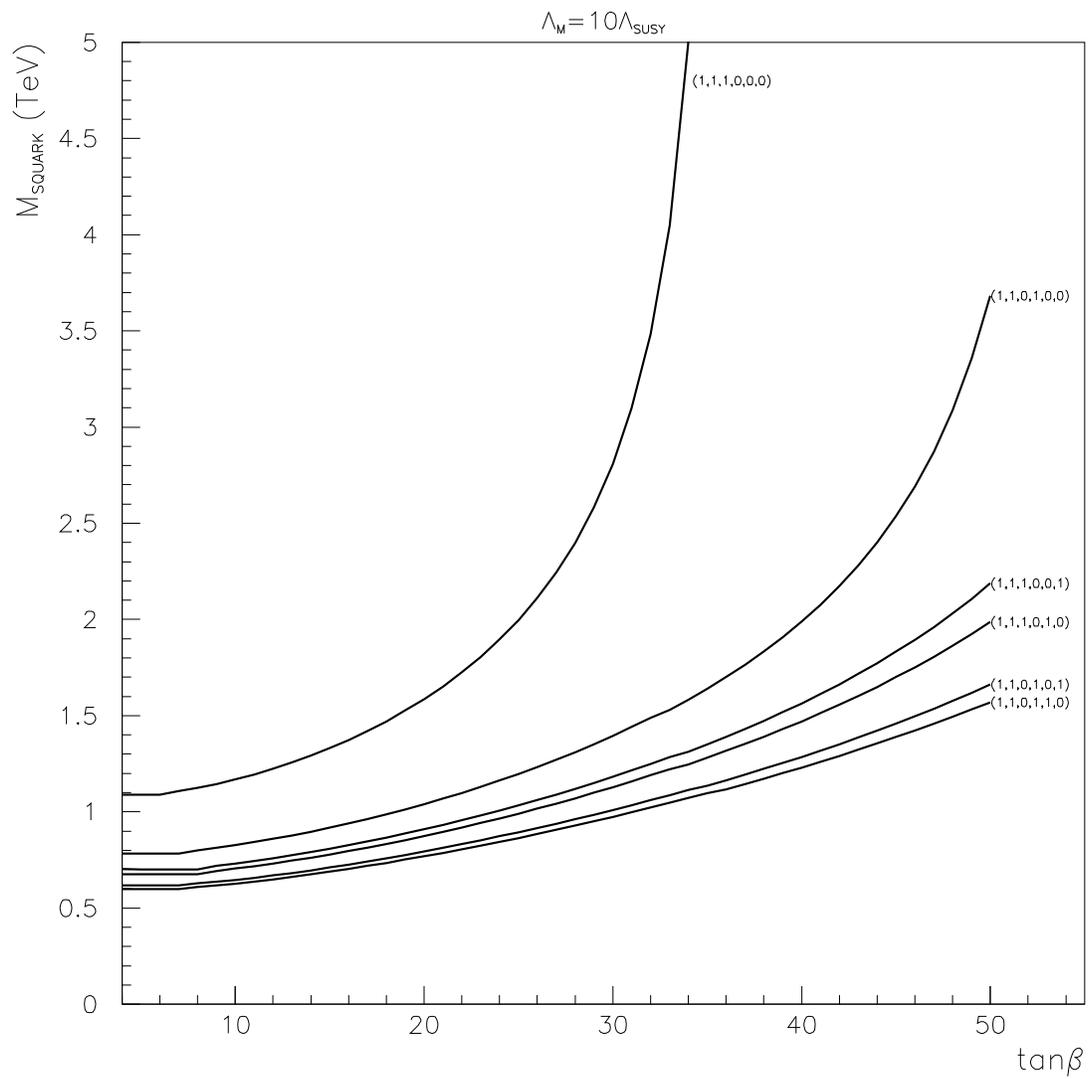
\end{document}